**NephroNet: A Novel Program for Identifying Renal Cell Carcinoma and Generating Synthetic Training Images with Convolutional Neural Networks and Diffusion Models**

Yashvir Sabharwal

Battlefield High School

**Author Note**

Yashvir Sabharwal  https://orcid.org/0000-0003-3550-0323

The author declares no conflicts of interest.

The author acknowledges:

Ezekiel Wotring, Stability AI, for providing knowledge and advice on artificial intelligence as well as providing access to commercial hardware.

**Abstract**

Renal cell carcinoma (RCC) is a type of cancer that originates in the kidneys and is the most common type of kidney cancer in adults. It can be classified into several subtypes, including clear cell RCC, papillary RCC, and chromophobe RCC. In this study, an artificial intelligence model was developed and trained for classifying different subtypes of RCC using ResNet-18, a convolutional neural network that has been widely used for image classification tasks. The model was trained on a dataset of RCC histopathology images, which consisted of digital images of RCC surgical resection slides that were annotated with the corresponding subtype labels. The performance of the trained model was evaluated using several metrics, including accuracy, precision, and recall, and achieved an overall accuracy of 96%. Additionally, in this research, a novel synthetic image generation tool, NephroNet, is developed on diffusion models that are used to generate original images of RCC surgical resection slides. Diffusion models are a class of generative models capable of synthesizing high-quality images from noise. Several diffusers such as Stable Diffusion, Dreambooth Text-to-Image, and Textual Inversion were trained on a dataset of RCC images and were used to generate a series of original images that resembled RCC surgical resection slides, all within the span of fewer than four seconds. The generated images were visually realistic and could be used for creating new training datasets, testing the performance of image analysis algorithms, and training medical professionals. NephroNet is provided as an open-source software package and contains files for data preprocessing, training, and visualization. Overall, this study demonstrates the potential of artificial intelligence and diffusion models for classifying and generating RCC images, respectively. These methods could be useful for improving the diagnosis and treatment of RCC and other types of cancer, as they provide a more efficient and accurate way of analyzing and generating images.

# Table of Contents

**Introduction**

Renal cell carcinoma (RCC) is a type of cancer that originates in the kidneys and is the most common type of kidney cancer in adults (Zhu et al., 2021). Classifying RCC by type and generating high-quality synthetic training images are new areas of particular interest and have applications from cancer prognosis and diagnosis to training medical professionals in nephrology. For instance, a whole-slide image of clear cell RCC can be classified and visualized, validating the inference of surgeons and technicians. Additionally, papillary RCC can be generated in a matter of seconds to demonstrate examples, contribute to a dataset, or be utilized as a training method. These opportunities present new ways to develop effective, accurate, efficient, and incredibly fast biotechnology.

Three main histologic RCC subtypes make up the RCC categorization. The most frequent subtype of renal cell carcinoma (about 75% of all instances) is clear cell renal cell carcinoma (ccRCC), followed by papillary renal cell carcinoma (pRCC), which accounts for about 15-20% of RCC, and chromophobe renal cell carcinoma (chRCC), which accounts for around 5% of RCC (Ricketts et al., 2018). Compact, alveolar, tubulocystic, or, less frequently, papillary architecture of cells with transparent cytoplasm and a distinctive network of tiny, thin-walled vessels are the typical morphologic characteristics of ccRCC (Udager & Mehra, 2016). PRCC is distinguished by the presence of papillae or tubulopapillary architecture with fibrovascular centers and often foamy macrophages (Valenca et al., 2015). It should be noted that although renal oncocytoma is the most prevalent kind of benign kidney tumor, it is still challenging to tell the two apart clinically from renal cell carcinoma, particularly chRCC (Dey et al., 2019).

Deep learning allows computational models that are composed of multiple processing layers to learn representations of data with multiple levels of abstraction (Lecun et al., 2015). Specifically, convolutional neural networks (CNNs) are primarily used to solve difficult image-driven pattern recognition tasks and with their precise yet simple architecture, offer a simplified method of getting started with artificial neural networks (ANNs), or biologically inspired computational models that can far exceed the performance of previous forms of artificial intelligence in common machine learning tasks (O'Shea & Nash, 2015). ResNet-18, a CNN, is designed to recognize objects in images. It is trained to classify images into one of 1000 different object categories, such as dog, cat, car, and so on. ResNet-18 is a smaller version of the ResNet network architecture, which was developed by researchers at Microsoft. The "18" in its name refers to the fact that it has 18 layers for data to process to. ResNet-18 is typically used for image classification tasks, and it has achieved state-of-the-art performance on many benchmarks. ResNet-18 has a structure similar to other CNNs, with an input layer, several hidden layers, and an output layer. The input layer takes in raw image data, and the hidden layers use convolutional, pooling, and activation functions to extract meaningful features from the images. These features are then used by the output layer to classify the images into the appropriate category.

In residual networks such as ResNet-18, the formulation of F(x) + x can be realized by feedforward neural networks with "shortcut connections", which are those skipping one or more layers (He et al., 2016). The entire network can still be trained end-to-end with backpropagation and can be easily implemented using common libraries without modifying the solvers (He et al., 2016). This is depicted in Figure 1.

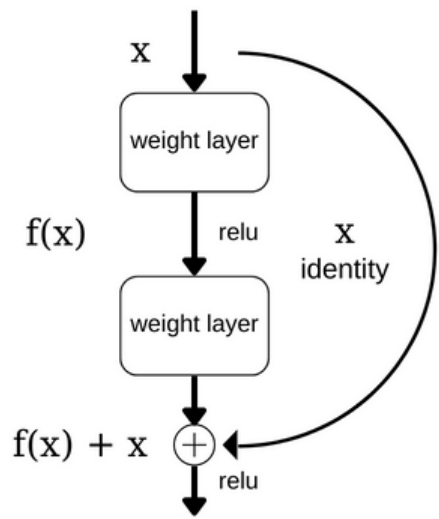

**Figure 1. A building block of residual learning.** The shortcut connections simply perform identity mapping, and their outputs are added to the outputs of the stacked layers (He et al., 2016).

In a study conducted by Zhu et al. (2021), a deep neural network artificial intelligence model was developed to accurately classify digital surgical resection slides and biopsy slides into five related RCC classes: clear cell RCC, papillary RCC, chromophobe RCC, renal oncocytoma, and normal or benign. Their whole-slide RCC classification pipeline was visualized for interpretation and provided identification to indicative regions by generating thousands of patches and performing grid-search technology. This general approach allows the pipeline to be modified and can support other sources of data and origin of specimen or specimen type. The reorganization and classification of complex histological patterns of RCC on biopsy and surgical resection slides under a microscope remains a heavily specialized, error-prone, and time-consuming task for pathologists and the intense effort or concentration required to successfully identify RCC classes can be relieved with these new approaches (Zhu et al., 2021).

In this research, two deep learning tools, a modified version of DeepSlide and NephroNet, are trained on a set of large, high-quality, and diverse microscopy images of kidney tissue to classify and generate RCC. This dataset is obtained from the Department of Pathology and Laboratory Medicine at Dartmouth-Hitchcock Medical Center (DHMC). The DHMC dataset includes 563 hematoxylin and eosin (H&E)-stained formalin-fixed paraffin-embedded (FFPE) whole-slide images of RCC (Zhu et al., 2021) that have been de-identified or anonymized before release. The images are labeled with the predominant histological pattern of each whole-slide image including renal oncocytoma, chRCC, ccRCC, and pRCC (Zhu et al., 2021). The DeepSlide tool uses the ResNet-18 residual network model and architecture for its classification abilities. NephroNet adopts latent diffusion models (LDMs) for their ability to achieve new state-of-the-art scores for image inpainting and class-conditional image synthesis and their highly competitive performance on various tasks, including text-to-image synthesis, unconditional image generation and super-resolution, while significantly reducing computational requirements compared to pixel-based diffusion models (Rombach et al., 2022).

Of the 563 formatted images, 233 were used for training the DeepSlide model, 110 were used for validating the model, and 220 were used for testing. The original code by Zhu et al. (2021) was modified to improve code with optimization, used newly updated libraries and frameworks, and utilized custom scripts for downloading, data preprocessing, directory management, and more. Results and metrics were taken from finding accuracy, time to use, and ease of access. NephroNet, the second and completely original tool developed in this study, used all 563 images from the DHMC RCC dataset to use for training. Results and metrics were taken from finding generation time, model training, and ease of access. NephroNet was compiled with four different diffusers (normal Stable Diffusion, Dreambooth Text-to-Image with modified UNet, Dreambooth Text-to-Image with modified UNet and text encoder, and Textual Inversion). As classifying RCC or generating RCC images specifically with artificial intelligence is a relatively new field, there are no competing tools or programs online to benchmark against. The original code for DeepSlide can be accessed at https://github.com/BMIRDS/deepslide. The NephroNet tool is open-source and can be accessed at https://doi.org/10.5281/zenodo.7498108.

## Materials & Methods

### Model Training Dataset

The model training dataset is derived from whole-slide images of surgical resections of kidney tissue by the DHMC. The dataset images were reviewed by a board and were released after they were anonymized. These slides focus on RCC and were labeled with their type by two pathologists at DHMC. The slides were originally scanned by an Aperio AT2 whole-slide scanner at 20x magnification and converted to Portable Network Graphics, or PNG format using libvips at 5x magnification (Zhu et al., 2021). After reviewing and sorting the slides, all of them were used as a model dataset. Approximately 41% were used for training (233), 20% for development or validation (110), and 39% for testing (220).

### Model Building (DeepSlide)

Classifying snapshots of renal cell carcinoma with the use of artificial intelligence can ease the effort, time, and knowledge required to successfully separate them manually. By using CNNs, we can use patch-prediction aggregation strategies to perform a grid search for feeding into the model. After the model runs, it has the capability to visualize the process occurring in the layers to a legible format for understanding. Patches were generated from the set of 563 RCC images in 224x224 pixel resolution. These were exported in JPG format and about 23,655 images were balanced per class (11 total classes) at a slide overlap of ½ to conserve system resources. White space in the histology images were automatically removed. There were 40 total epochs with a learning rate of 0.001 and a weight decay of 0.0001. The learning rate decayed by 0.9 every epoch.

The DeepSlide code was modified from the original program by using new packages, assigning different hyperparameters, creating runtime scripts, and optimizing code. The architecture consisted of an 18-layer convolutional neural network. Data is inputted from the first layer, Input, and presents the final classification through the last layer, Output. Similar to other CNNs, the model is a 3-dimensional array with the dimensions of width, height, and channel, where channel represents the number of color channels (3 channels for RGB images). The model was trained in PyTorch 3.19 on a RTX A5000 graphics card. Total training time for 40 epochs was 5 hours and 51 minutes.

**Model Building (NephroNet)**

Generating synthetic images with artificial intelligence provides opportunities for medical advancement in doctor training, data creation, benchmarking, and much more. By using diffusers, we can modify the original Stable Diffusion model provided to us. Generally speaking, diffusion models are machine learning systems that are trained to denoise random Gaussian noise step by step, to get to a sample of interest, such as an image (Patil et al., 2022). This method is accompanied by transforming attributes in latent space to reduce memory usage. A logical workflow of Stable Diffusion can be seen in Figure 3. During the study, 3 different diffusers along with the original Stable Diffusion model were used to find the best results. These diffusers were Dreambooth with finetuned UNet (Version 1), Dreambooth with finetuned UNet and Text Encoder (Version 2), and Textual Inversion (Version 3), respective to the order the models were trained in. As a control group or testing base, Stable Diffusion version 1.5 with no modification (Control Group) was used. All diffusers used the DHMC RCC dataset, and the input images were converted to 512x512 resolution. Every trial used the same prompt, "a photo of renal cell carcinoma", and the same number of steps, 50.

All development was conducted in PyTorch 3.19 on a RTX A5000 graphics card. The software package, NephroNet, contains all of the models, runtime scripts, and results gathered from the experiment. The Control Group was previously trained and runs from a pool of GPU's as it is a commercial application and is the only application not developed by the researcher. Total generation time was approximately 10-15 seconds per image.

Dreambooth presents a new approach for "personalization" of text-to-image diffusion models (adapting them to user-specific image generation needs) (Ruiz et al., 2022). With Version 1, the Dreambooth pipeline was utilized with model weights from CompVis Stable Diffusion version 1.4. This pipeline took in the input dataset and a prompt to output an image similar but completely original to the RCC images. The input images is what modifies the UNet. Hyperparameters for Version 1 can be viewed in Table 1. Version 2 used the exact same pipeline with the only difference being an extra hyperparameter of *train text encoder*, which modifies the Text Encoder as well as the UNet. Model training for Version 1 took 6 minutes and 3 seconds.

Model training for Version 2 took 6 minutes and 27 seconds. Both versions took 4 seconds per image to generate.

Textual Inversion transforms the world of image generation by finding new words in the textual embedding space of pre-trained text-to-image models (Gal et al., 2022). An input string is first converted to a set of tokens, where each token is then replaced with its own embedding vector, and these vectors are fed through the downstream model (Gal et al., 2022). With Version 3, the Textual Inversion pipeline was used with the same CompVis model as Versions 1 and 2. Textual Inversion introduces new hyperparameters for passing the token. First, a *learnable property* is given, such as "modality" for image or "body part" for kidney. Next, a *placeholder token* and *initializer token* are passed. For these, "<kidney cancer-image>" and "image" were passed, respectively. For this model's training, 3000 steps were cycled through, and it took a total of 57 minutes and 34 seconds. Version 3 took 3 seconds to generate an image. The usage and model building workflows can be viewed in Figure 2.

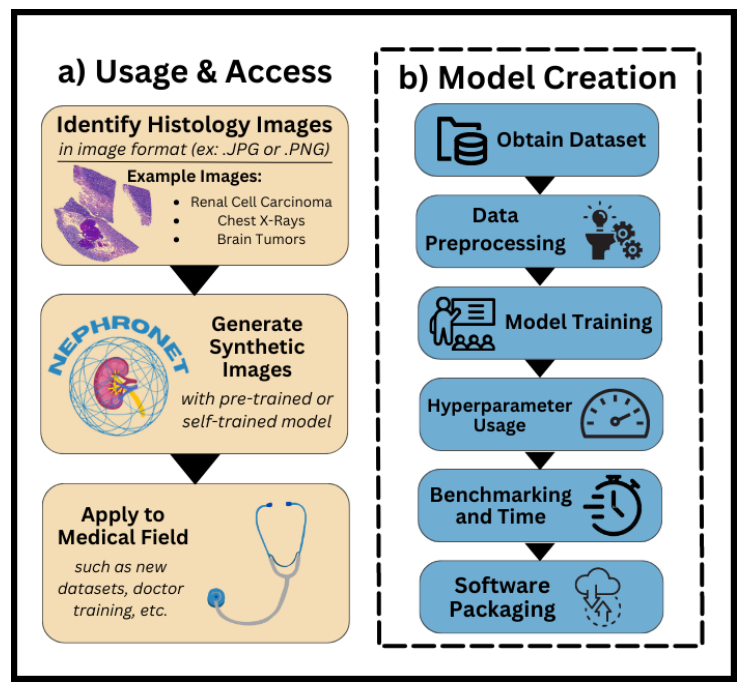

**Figure 2.** **Usage and model creation workflows for RCC synthetic image generation using NephroNet deep learning models and software package.**

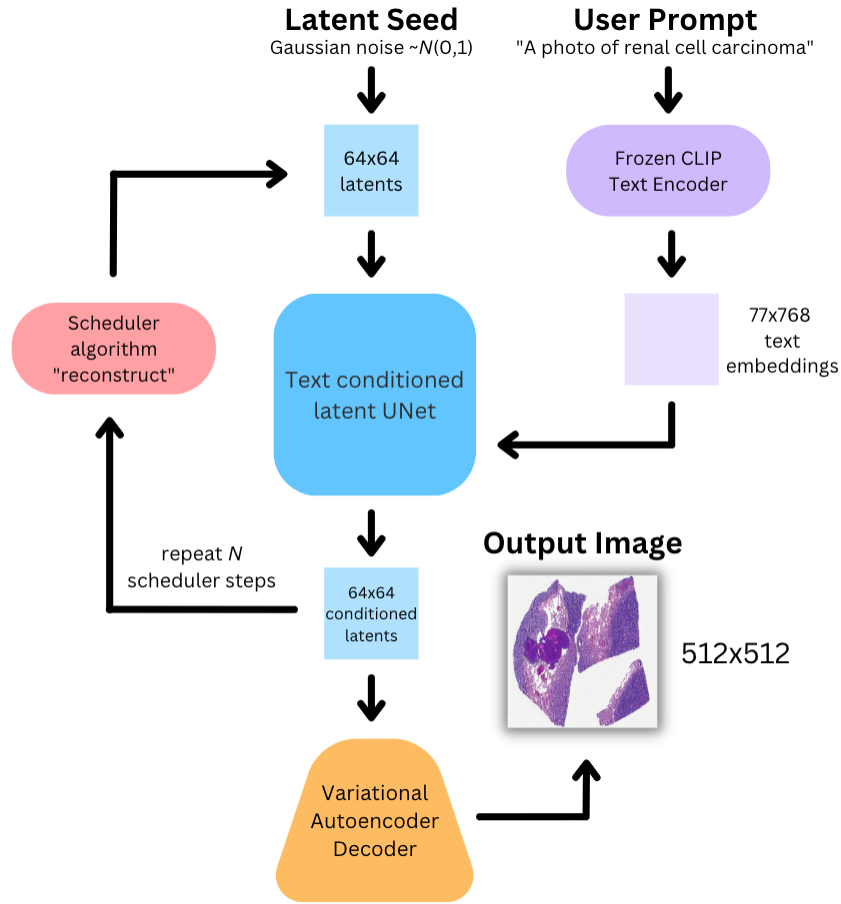

**Figure 3. Diagram of logical workflow of latent diffusion models.** This diagram represents a logical explanation of latent diffusion models such as Stable Diffusion. First, a seed is generated and noise is added to the latents. These latents are fed into the UNet along with text embeddings from the user prompt that has been run through a frozen CLIP text encoder. These are turned into conditioned latents and is repeated for the number of scheduler steps. The scheduler adjusts learning rates as defined and the process is repeated. Conditioned latents are then sent through the variational autoencoder decoder to output the desired image in a 512x512 pixel resolution.

**Hyperparameter Tuning**

There were several variables experimented with model building and training such as batch sizes, learning rates, and training steps as displayed in Table 1.

Hyperparameters are parameters that are set before training a machine learning model and are not learned during training. They are used to control the learning process and the behavior of the model. They are important in model building because they have a significant impact on the model's performance and can significantly affect the model's ability to generalize to new data. For example, the learning rate, which determines the step size at which the model updates its weights during training, is a hyperparameter. If the learning rate is too high, the model may overshoot the optimal weights and never converge to a good solution. On the other hand, if the learning rate is too low, the model may take a long time to train and may not achieve a high level of accuracy.

| Hyperparameter | Value |
|---|---|
| Model Name (the pre-trained Stable Diffusion model to gather weights from) | CompVis/stable-diffusion-v1-4 |
| Instance Prompt (the phrase that the model will train off the dataset from) | "a photo of renal cell carcinoma" |
| Resolution (the resolution of the image produced) | 512 x 512 pixels (512) |
| Train Batch Size (number of samples propagated) | 1 |
| Gradient Accumulation Steps (split the batch) | 1 |
| Learning Rate (controls the weights and pace at which the algorithm learns or updates) | 2e-6 |
| Learning Rate Scheduler (a framework that controls the learning rate between epochs) | constant |
| Maximum Training Steps (how many steps the model will go through) | 650 |

**Table 1. Hyperparameter variables are utilized to achieve the best outcomes.** (A) On the left-hand column, 8 hyperparameters are listed. (B) On the right-hand column, values to the specified hyperparameters are listed. Hyperparameters are used to find the best settings for model training and receive the best results. These are also used to prevent over/underfitting.

**Software Architecture**

The development of the modified DeepSlide application and NephroNet have miltiple software components that make up the workflow for the end user. These include sequential steps for DeepSlide, NephroNet's customized model architectures, and several runtime scripts that help preprocess data, run the model trainings, and generate multiple images at once.

Development steps for DeepSlide are as follows:

1. Data preprocessing takes all the images from the provided dataset and classifies them into three different classes: training, validation, and testing.
2. Each image in every class has several 224x224 patches generated.
3. The model is trained on the dataset.
4. The model is run on the testing data and patches.
5. A grid search is performed to find the best thresholds to throw out noise.
6. Predictions made by the model are visualized. An example can be seen in Results.
7. Final testing occurs and accuracy is provided.

Development steps for NephroNet are as follows:

1. Data preprocessing formats all the images from the provided dataset into 512x512 resolution. Each image is placed under one root directory.
2. Runtime scripts initiate model training and specify hyperparameters.
3. Inference pipelines are customized and run to generate a specified number of images with a prompt on the trained model.

As everything is developed with PyTorch, an open-source artificial intelligence framework, modifying and improving both DeepSlide and NephroNet are easy for the end user. Models can be packaged later for compatibility with other development services such as MATLAB.

**Hardware Requirements**

The hardware requirements for training a machine learning model depend on a number of factors, including the size and complexity of the model, the amount of data being used, and the desired training speed. As the dataset images are very high-quality, the archive files containing them amount to a large size that requires hardware capable of performing to industry standards. In this case, using consumer hardware is inefficient and more costly than renting enterprise solutions on the cloud. In this research, a server in Great Britain was rented for approximately $0.44 USD per hour. This training cost is relatively low and allows models or inferences to be run faster. The server's main compute came from an RTX A5000 graphics card and the server was deployed on a secure cloud which allows you to reserve a machine for uninterrupted access. The operating system was Ubuntu 20.04 LTS. To save as many resources as possible, code development was conducted on a personal laptop and later uploaded to the server for execution. The total cost for all research was approximately $20 USD.

**Statistical Analysis**

Accuracy for the model and generation time for images were used as the primary metrics to conduct statistical analysis in the modified DeepSlide program and NephroNet, respectively. Accuracy is a common metric used to evaluate the performance of a machine learning model. It is the fraction of predictions made by the model that are correct. To determine the accuracy of a model, we have a dataset with known, correct labels, which will be used as a reference to compare the model's predictions against. These datasets are typically split into two parts: a training set and a test set, but for research purposes, we divided them into three parts: training, validation, and testing. The model is first trained on the training set and then its accuracy is evaluated on the test set. The generation time for a diffusion model is simply the average time it takes for a new image to be produced in the pipeline. As there are no competing tools, programs, or benchmarks to compare against, these results provide a baseline for further testing and research.

## Results

A few primary metrics are used to determine the performance of the modified DeepSlide work and NephroNet. These include accuracy (DeepSlide), model training time (DeepSlide & NephroNet), and general visualization (NephroNet).

### Accuracy (DeepSlide)

Based on the 563 images referenced from the DHMC histopathology dataset, the model trained received a total accuracy of 0.96, or 96%. Due to time restraints, F1-score, recall, or precision were not recorded. An accuracy of 96% is generally considered to be very good performance for a machine learning model. It means that out of all the predictions made by the model, 96% of them were correct. A demonstration of how patches were removed can be seen in Figure 4.

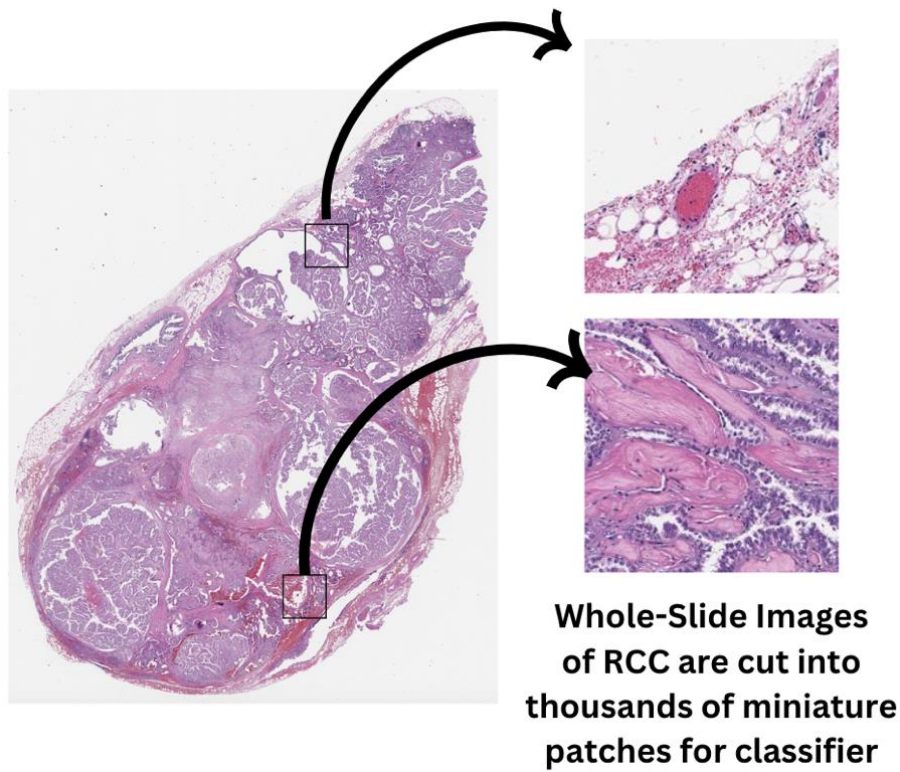

**Whole-Slide Images of RCC are cut into thousands of miniature patches for classifier**

**Figure 4. A representation of patch generation from a whole-slide image of RCC.** Several extremely small images are taken from high-resolution images for model classifying and training. These images are 512x512 pixels.

**Training Time (DeepSlide & NephroNet)**

Training time is an important factor to consider when building machine learning models, especially when the model is large and complex. It can take a significant amount of time to train a machine learning model, and this time can vary greatly depending on a number of factors. Some factors that can affect training time include model size and complexity, data size, hardware, and hyperparameters. Larger and more complex models tend to take longer to train, as they have more parameters that need to be learned. Training time can also be affected by the amount of data being used. In general, models trained on larger datasets will take longer to train. The hardware being used for training may also have a significant impact on training time. For example, training a model on a graphics card can be much faster than training on a CPU. The hyperparameters of the model, such as the learning rate and the batch size, might affect training time. In general, it is important to consider training time when building machine learning models, and to choose hardware and hyperparameters that are appropriate for the task at hand.

For DeepSlide, development time took several days, but the model training itself took a total of 5 hours and 51 minutes. The model was trained on 40 epochs and more information can be found in the Model Building section. A variety of times were logged for the development of NephroNet. Data is included in Table 2 for the different diffusers, training time, generation time, and more for Stable Diffusion, Dreambooth, and Textual Inversion.

|  | Training Time | Generation Time (per image) | Real-World Usage | Modifiers |
|---|---|---|---|---|
| **Control Group** | N/A | 10-15 seconds | Bad (3) | N/A |
| **Version One** | 00:06:03 | 4 seconds | Best (1) | UNet |
| **Version Two** | 00:06:27 | 4 seconds | Good (2) | UNet & Text Encoder |
| **Version Three** | 00:57:34 | 3 seconds | Worst (4) | Tokenizer |

**Table 2. Data for various statistics including model training time.** On the left-hand side, the various models used for image generation are listed. On the top, different values for performance gauging are listed. In the first column, the training time for how long the model took to train is stated. The next column represents how long the program took to generate a single image. The third column shows opinions on effectiveness in the real world (1 being the best, 4 being the worst). The last column shows which modifiers were taken into consideration during training. The Control Group and Versions 1, 2, and 3 can be read in-depth above under Materials & Methods.

**General Visualization (NephroNet)**

Image generation using artificial intelligence is an active area of research, and new approaches and techniques are being developed all the time. It is an exciting area with many potential applications, and it has the potential to revolutionize how we create and use images. With NephroNet, medical professionals can use the program to generate synthetic images for training doctors, create new datasets for learning, find an alternative to releasing patient data requiring consent, use as a benchmark for other models in the future, and skip the process of going through an entire surgical procedure to get one slide from one type of kidney cancer. As an overview, NephroNet can help with the diagnosis, treatment, and growth of knowledge for RCC.

The images by the diffusion models are generated in just seconds with the prompt "a photo of renal cell carcinoma". With more testing, better results can be achieved. A figure of output images for each version and the control group can be seen in Figure 5.

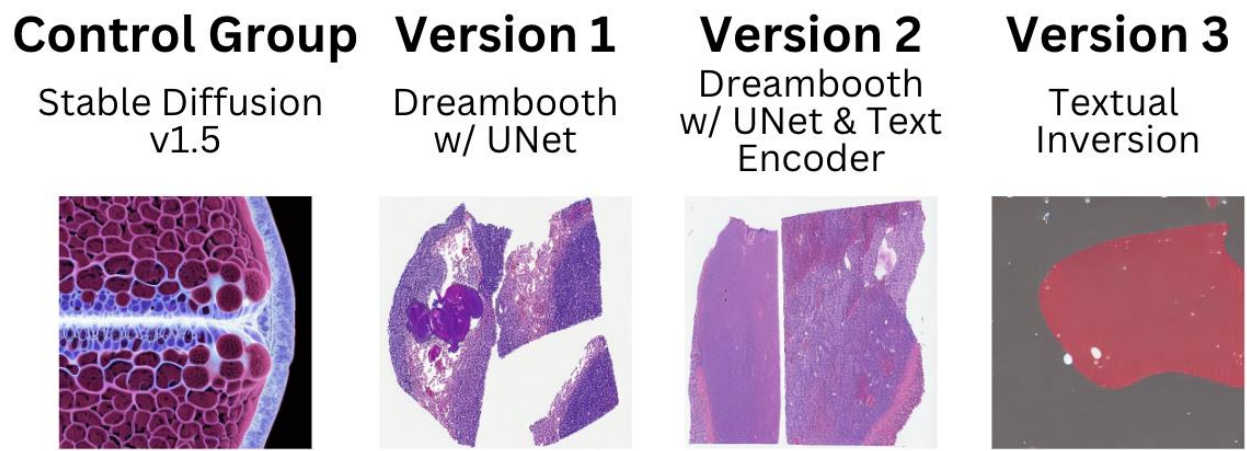

**Figure 5. Results from diffusion models.** From left to right, the Control Group and specific versions of testing are listed in order. Below the titles, the description of the models is stated. Each column has an image of the output from generating with the prompt "a photo of renal cell carcinoma". The Control Group generated an image that looks most like "clipart" or a generic illustration of what RCC may represent. Version 1 achieved the best results and shows the whole-slide image in full detailed with correct staining. Version 2 had an effort to try and replicate the input images as best as possible but lacked in retaining fine-point quality. Version 3 output the worst image of the tests; this may be due to underfitting or overfitting.

## Discussion

In this research study, two novel artificial intelligence programs are proposed that use convolutional neural networks and diffusion models to both classify and generate renal cell carcinoma. It is found that machine and deep learning are particularly effective methods in cancer classification and synthetic image generation because they can use trained models to classify images by type of abnormality and create completely original visualizations of whole-slide images containing surgical resections of RCC. As there are no previous works or studies using the method of NephroNet, findings in this research provide baseline benchmarks for future development and exploration in the biomedical engineering domain. Using these approaches, the modified DeepSlide program and NephroNet architecture were built upon a high-quality dataset of whole-slide images from the DHMC. Having a dataset that includes multiple examples of several types of RCC as well as the labels to join them with increases accuracy and the ability to effectively learn exponentially. Using variational autoencoder decoders and transformers also aided in reducing the amount of compute required to handle these large processes.

This research also demonstrates that generating synthetic images with diffusers for Stable Diffusion as a technique is of particular interest. The Dreambooth diffuser with a modified UNet is utilized in the final NephroNet model. Other diffusers, such as Stable Diffusion v1.5, Dreambooth with a modified UNet and text encoder, and Textual Inversion were also used to determine best results. After successful testing, Version 1 of the research provided the fastest training time, most accurate representation, and swift generation periods. Additional inferencing pipelines were created to generate multiple images with different prompts or passing specific tokens and modalities to the machine. Overall, this new and developing technologies has the opportunity to break medical barriers with further testing.

The statistical data analysis uses accuracy as a way to directly compare the original DeepSlide code against the modified DeepSlide program in this study. Although accuracy was 0.01 or 1% lower, the preprocessing and user ease of access was improved. For NephroNet, model training time, image generation periods, modifiers, and real-world comprehensiveness were all factored into consideration. Recent research has shown that creating models for synthetic image generation is possible, and such research can be applied to the results and future development of this study. Accuracy, field testing, approval of pathologists, and other metrics may provide valuable insight into the results of the tools.

In research for NephroNet, the study's main focal point, four diffusion model generation options were evaluated. Starting with the Control Group, or Stable Diffusion v1.5, the LAION-5B large-scale dataset pre-trained model achieved mediocre results that did not align with the initial goals of the study. This model output generic images and drawings that referenced RCC but never accurately provided a detailed representation of a whole-slide image and image generation time took anywhere from ten to fifteen seconds. This conclusion ranked third out of the four. The next model, Version One, consisted of a Dreambooth diffuser that modified the UNet of the model by inputting provided dataset images. Model training was fast for industry standards and image generation yielded an average of four seconds. This model output accurate images that referenced and represented whole-slide images of RCC. The positive results ranked this version first out of the four. The third model, Version Two, was similar to the first with the only difference being an extra hyperparameter that trained and provided a modified text encoder that replaced the frozen CLIP text encoder. Model training time was similar to the first and image generation yielded an average of four seconds per image. Version Two ranked second out of the four. The last model, Version Three, is based on Textual Inversion. By using dataset images and tokens as input, the model generates images based off of your prompt. As the name of the model states, most modification goes into the text portion. Unfortunately, this model resulted in under or overfitting and produced the worst images from the testing group. Model training time took fifty-seven minutes and thirty-four seconds. Image generation took about three seconds per image. Version Three ranked fourth out of the four models.

NephroNet synthetic image generation is competitive and can be applied directly in diagnosis and treatment, doctor training, dataset creation, and several more applications. Currently, this innovative technology can be accessed through an open-source software package. Although NephroNet is based on renal cell carcinoma, it may be plausible to apply this methodology to other areas such as chest x-rays in future research. With more time, resources, and people, NephroNet can be heavily improved. Finally, future study may consider testing NephroNet in the lab environment and using more powerful hardware to further confirm validity in the results.

## Bibliography

Dey, S., Noyes, S. L., Uddin, G., & Lane, B. R. (2019). Palpable Abdominal Mass is a Renal Oncocytoma: Not All Large Renal Masses are Malignant. *Case Reports in Urology*, *2019*, 1–4. https://doi.org/10.1155/2019/6016870

Gal, R., Alaluf, Y., Atzmon, Y., Patashnik, O., Bermano, A. H., Chechik, G., & Cohen-Or, D. (2022). *An Image is Worth One Word: Personalizing Text-to-Image Generation using Textual Inversion*. https://doi.org/10.48550/arxiv.2208.01618

He, K., Zhang, X., Ren, S., & Sun, J. (2016). *Deep Residual Learning for Image Recognition* (pp. 770–778). http://image-net.org/challenges/LSVRC/2015/

Lecun, Y., Bengio, Y., & Hinton, G. (2015). Deep learning. *Nature 2015 521:7553*, *521*(7553), 436–444. https://doi.org/10.1038/nature14539

O'Shea, K., & Nash, R. (2015). *An Introduction to Convolutional Neural Networks*. https://doi.org/10.48550/arxiv.1511.08458

Patil, S., Cuenca, P., Lambert, N., & von Platen, P. (2022). *Stable Diffusion with  Diffusers*. https://huggingface.co/blog/stable_diffusion

Ricketts, C. J., de Cubas, A. A., Fan, H., Smith, C. C., Lang, M., Reznik, E., Bowlby, R., Gibb, E. A., Akbani, R., Beroukhim, R., Bottaro, D. P., Choueiri, T. K., Gibbs, R. A., Godwin, A. K., Haake, S., Hakimi, A. A., Henske, E. P., Hsieh, J. J., Ho, T. H., … Mariamidze, A. (2018). Erratum: The Cancer Genome Atlas Comprehensive Molecular Characterization of Renal Cell Carcinoma (Cell Reports (2018) 23(1) (313–326.e5) (S2211124718304364) (10.1016/j.celrep.2018.03.075)). *Cell Reports*, *23*(12), 3698. https://doi.org/10.1016/j.celrep.2018.06.032

Rombach, R., Blattmann, A., Lorenz, D., Esser, P., & Ommer, B. (2022). *High-Resolution Image Synthesis with Latent Diffusion Models*. 10674–10685. https://doi.org/10.1109/CVPR52688.2022.01042

Ruiz, N., Li, Y., Jampani, V., Pritch, Y., Rubinstein, M., & Aberman, K. (2022). *DreamBooth: Fine Tuning Text-to-Image Diffusion Models for Subject-Driven Generation*. https://doi.org/10.48550/arxiv.2208.12242

Udager, A. M., & Mehra, R. (2016). Morphologic, Molecular, and Taxonomic Evolution of Renal Cell Carcinoma: A Conceptual Perspective With Emphasis on Updates to the 2016 World Health Organization Classification. *Archives of Pathology & Laboratory Medicine*, *140*(10), 1026–1037. https://doi.org/10.5858/ARPA.2016-0218-RA

Valenca, L. B., Hirsch, M. S., Choueiri, T. K., & Harshman, L. C. (2015). Non-Clear Cell Renal Cell Carcinoma, Part 1: Histology. *Clinical Advances in Hematology & Oncology*, *13*(5).

Zhu, M., Ren, B., Richards, R., Suriawinata, M., Tomita, N., & Hassanpour, S. (2021). Development and evaluation of a deep neural network for histologic classification of renal cell carcinoma on biopsy and surgical resection slides. *Scientific Reports 2021 11:1*, *11*(1), 1–9. https://doi.org/10.1038/s41598-021-86540-4